\documentclass[pra,reprint,superscriptaddress,showpacs,amsmath, floatfix,amssymb,aps]{revtex4-1}
\usepackage{graphicx}
\usepackage{subfigure}
\usepackage{dcolumn}
\usepackage{bm}
\usepackage{hyperref}
\usepackage{indentfirst}
\usepackage{braket}
\usepackage{amsmath}
\usepackage{ulem}
\usepackage{color}
\hypersetup{colorlinks=true, citecolor=blue, urlcolor=blue, linkcolor=blue}
\begin{document}

\title{Phase diagram of the two-component bosonic system with pair hopping in synthetic dimension }
\author{Chenrong Liu}
\email{crliu@wzu.edu.cn}
\affiliation{College of Mathematics and Physics, Wenzhou University, Zhejiang 325035, China}

\author{Zhi Lin}
\email{zhilin18@ahu.edu.cn}
\affiliation{School of Physics and Materials Science, Anhui University, Hefei 230601, China}
\affiliation{State Key Laboratory of Surface Physics and Department of Physics, Fudan University, Shanghai 200433, China}
\date{\today}
\begin{abstract}
We systematically study the ground-state phase diagrams and the demixing effect of a two-dimensional two-component bosonic system with pair hopping in synthetic dimensions by using the cluster Gutzwiller mean-field method. Our results show that when the interexchange symmetry between the two species is broken, the regions of the super-counter-fluidity state in the phase diagram are dramatically shrunk whenever the on-site pair hopping term is turned on or off. Unexpectedly, the non-integer Mott phase and the molecular superfluid phase predicted in our previous work \citetext{Z. Lin et al., \textcolor{blue}{Phys. Rev. Lett. \textbf{125}, 245301 (2020)}}, can only exist in such a system that the pair hopping term is opened, and more importantly, its interexchange symmetry must be broken. Moreover, the demixing effect of the two-component bosonic system with synthetic pair hopping interaction has also been studied, and we find that an incompletely demixed state is formed in the system when the strength of the interspecies on-site repulsive interaction is sufficiently large.
\end{abstract}
\maketitle

\section{Introduction}
Ultracold atomic system in the optical lattice opens a new window to the research of modern condensed matter physics \cite{RevModPhys71.463,AdvPhys56.243,Ultracold Atoms in Optical Lattices,RepProgPhys82.104401}.  The remarkable control of the interactions between the atoms leads to the experimental realization of several phase transitions, i.e. superfluid (SF) to Mott-insulator (MI) transition in a bosonic system \cite{Nature415}. Recently, more and more attentions have been transferred from the single-component bosonic systems to the two-component bosonic systems because of the successful realization of these systems in optical lattices \cite{Nature425,PhysRevA77.011603} and the deepening of theoretical studies \cite{PhysRevLett90.100401,Demler,PhysRevLett92.050402}.  It can be described by the two-component Bose-Hubbard model, which includes the intraspecies on-site interaction along with the interspecies on-site interaction, and gives rise to many new phases such as super-counter-fluidity state (SCF), paired superfluid state (PSF), two-component Mott insulator phase (2MI), and two-component superfluid phase (2SF) \cite{PhysRevLett90.100401,PhysRevLett92.050402}. In these studies, the interexchange symmetry ($a$ component $\leftrightarrow$ $b$ component) can be assumed,  implying the same hopping amplitude and the same intraspecies on-site interaction for each species. By breaking this symmetry, there should have more complex phase diagrams.

On the other hand,  time, momentum space, or internal states (the internal atomic degrees of
freedom, e.g., pseudospin) can also be considered as the synthetic dimensions \cite{PhysRevLett112.043001,Sdimensions}. In contrast to manipulating the single-particle hopping processes \cite{s-hopping1,s-hopping2,s-hopping3,s-hopping6,s-hopping7,s-hopping8}, we have proposed a feasible scheme
for manipulating  two-particle hopping process along a synthetic dimension (or synthetic pair hopping) in a two-component bosonic system  \cite{PhysRevLett125.245301} via the Floquet engineering \cite{Bukov,Goldman,Eckardt_eff,Weitenberg}. By adding this synthetic pair hopping (SPH) interaction to the two-component Bose-Hubbard model, two novel phases are revealed in our previous results \cite{PhysRevLett125.245301}: one is the non-integer Mott insulator (NMI) phase and the other is the molecular superfluid (MSF) phase. Besides, we know that for the phase diagram of a two-component bosonic system without SPH interaction, there are some significant differences between the interexchange symmetric case and interexchange asymmetric case. Therefore, we can infer that when the SPH interaction is considered, there are still some significant differences in the phase diagrams of the above two cases. Although the phase diagram of two-component bosonic systems with SPH interaction has been revealed in the interexchange asymmetric case \cite{PhysRevLett125.245301}, it is interesting to study the phase diagram of the system with interexchange symmetry.

Furthermore, a sufficiently strong interspecies repulsion can localize the two species in separated domains, and this leads to the broken of the spatial symmetry of the system. People have researched the critical condition of the spatial phase separation in a two-species Bose-Hubbard Hamiltonian without the SPH interaction, and they have shown that $U_{ab}/U=1$ ($U_{ab}$ is the interspecies on-site interaction and $U_{aa}=U_{bb}=U$ is the intraspecies on-site interaction ) is the critical value \cite{PhysRevA76.013604, JPhysSocJpn81.024001, PhysRevA89.057601, PhysRevA92.053610}.  When $U_{ab}$ is greater than $U$, the system will undergo a phase transition to the demixed state.  However, the study of this demixing effect in a two-component bosonic system with SPH interaction is still lacking. Therefore, we will discuss it by including the interexchange symmetry in this paper.

We have organized the remaining part in the following way. In Sec.~\ref{Model}, we present our theoretical model and the method of calculations. We also show the numerical results and the detailed analysis in Sec.~\ref{Results}. At last, we conclude our findings in Sec.~\ref{Conclusions}.

\section{Model and method}\label{Model}
The Hamiltonian of the two-component bosonic system with pair hopping in synthetic dimension can be read as \cite{PhysRevLett125.245301},
\begin{eqnarray} \label{Ham}
H &=-J \displaystyle\sum_{\langle i,j\rangle}\left(a_{i}^{\dagger} a_{j}+b_{i}^{\dagger} b_{j}+\text {H. c.}\right)-\mu \sum_{i}\left(n_{i,a}+n_{i,b}\right) \nonumber\\ &+\displaystyle\frac{U_{aa}}{2} \sum_{i} n_{i,a}\left(n_{i,a}-1\right)+\frac{U_{bb}}{2} \sum_{i} n_{i,b}\left(n_{i,b}-1\right) \nonumber\\&+U_{ab}\displaystyle\sum_{i} n_{i,a} n_{i,b}+W \sum_{i}\left(a_{i}^{\dagger} b_{i} a_{i}^{\dagger} b_{i}+ \text {H. c.}\right),
 \end{eqnarray}
 where $a_i$ ($b_i$) is $a$ ($b$) component boson annihilation operator and $n_{i,a}$ ($n_{i,b}$) is $a$ ($b$) component boson number operator. This Hamiltonian has also been proposed in a multi-band Bose system\cite{,PhysRevB83.195106,PhysRevLett111.205302}. Here, the first five terms describe the two component Bose-Hubbard model \cite{PhysRevLett90.100401,NewJPhys5.113,PhysRevLett92.050402} and the $W$ term represents the SPH interaction.  The interspecies interaction $U_{ab}$ is set to be repulsive which means $U_{ab} >0$. Moreover, the interexchange symmetry requests $U_{aa}=U_{bb}$ and it can be broken by tuning $U_{aa}$ and $U_{bb}$ to be not equal. The phase diagram of this model has been researched in our early work when the interexchange symmetry is broken \cite{PhysRevLett125.245301}. In this paper, one of our main goals is to obtain the ground-state phase diagrams for the interexchange symmetry is presented, and the $W$ term can be turned on or off.

\begin{figure}[htb]
\centering
\includegraphics[width=0.45\textwidth]{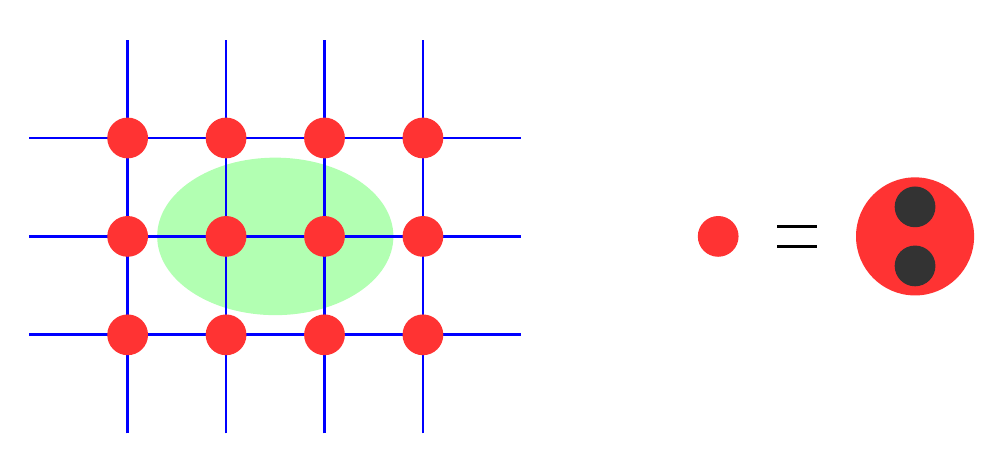}
\caption{\label{Fig1}(Color online) A illustration of the cluster which is used in our calculations. Each site in the two-dimensional square lattice is equivalent to two sub-sites in synthetic space.}
\end{figure}
To study the quantum phases of this model, the cluster Gutzwiller mean-field method \cite{PhysRevB86.144527,PhysRevA87.043619,JPhysB51.145302}, which can well capture the quantum fluctuations, is used to do these calculations. We choose the size of the cluster as $1\times 2$ which is equivalent to four sub-sites in synthetic space. The illustration is presented in Fig.\ref{Fig1}. For each component, the maximum occupation number per site is set up to 13 on account of the on-site repulsive interaction $U_{aa}$ and $U_{bb}$. Under these settings, the mean-field Hamiltonian matrix dimension is up to $D=14^4=38416$. The converge precision $\eta$ is set to $10^{-6}$ and the converge condition is $\Delta<\eta$, where $\Delta$ is defined as follows,
\begin{eqnarray}
\Delta=|E_{i}-E_{i-1}|+|\langle a \rangle_{i}-\langle a \rangle_{i-1}|+|\langle b \rangle_{i}-\langle b \rangle_{i-1}|.
\end{eqnarray}
which is the energy and order parameter difference between two continuous self-consistent steps. Thus, the energy and the single component SF order parameter can be both converged after the self-consistent process.

\section{Numerical results} \label{Results}
\subsection{$W=0$ case}\label{W=0}
\begin{figure*}[htb]
\centering
\includegraphics[width=0.9\textwidth]{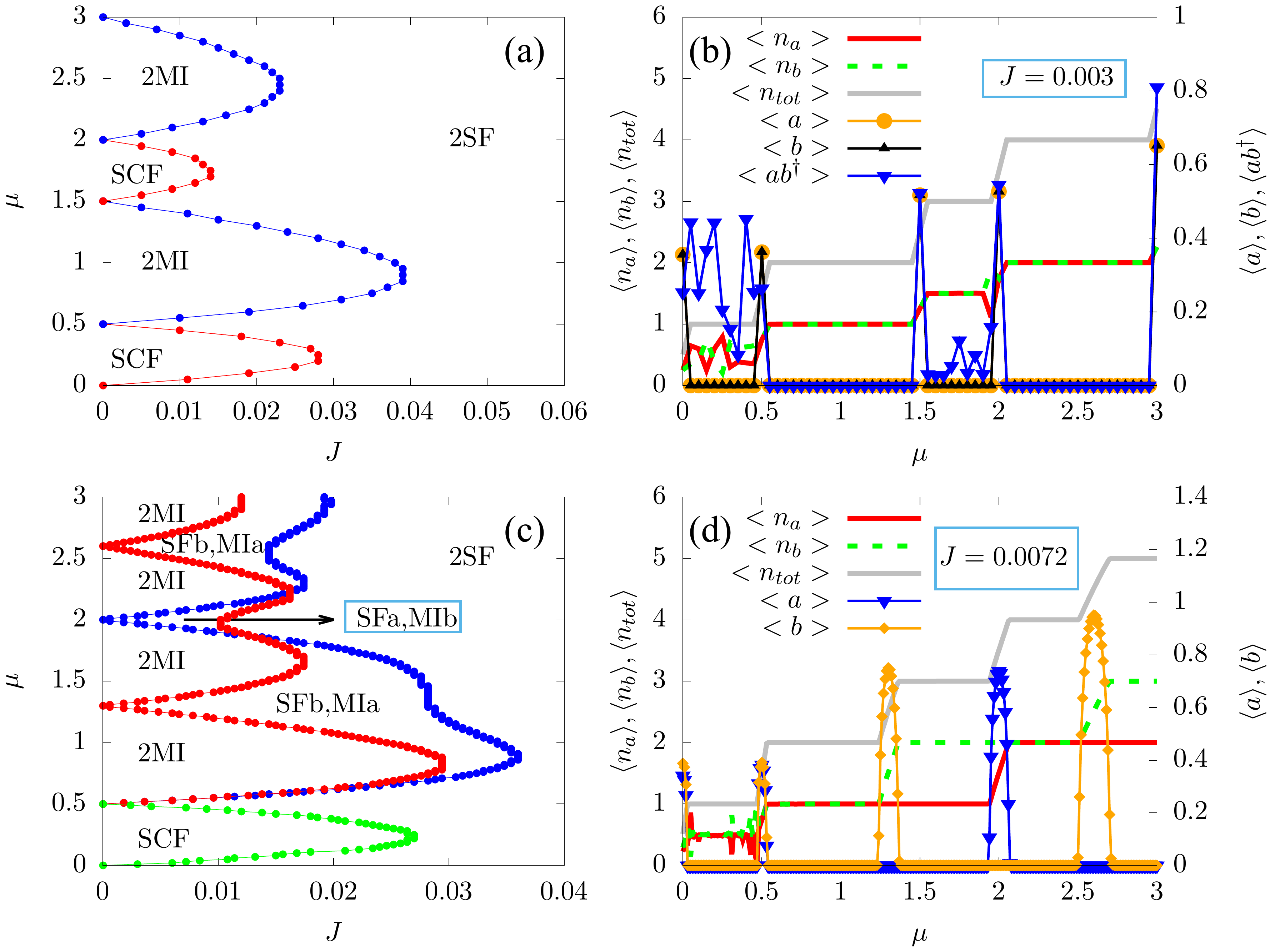}
\caption{\label{Fig2}(Color online) Chemical potential $\mu$ versus $J$ (both are in units of $U$) phase diagram for (a) $U_{bb}=U_{aa}=U$ and (c) $U_{bb}=0.8U_{aa}=0.8U$, here $U_{ab}$ is fixed at $0.5U$ and $W=0$.  Their corresponding physical measurements as a function of $\mu$ are also shown in (b) and (d) respectively at a fixed value of $J$. In (b) and (d), the left (right) vertical axis represents the value of the particle numbers (SF order parameters).}
\end{figure*}

The phase diagram of the two-component Bose-Hubbard model ($W=0$)  has been widely studied in two dimensions\cite{Iskin, PRA81.053608, arXiv2021}. For comparison, we reproduce the results in this part with $U_{aa}=U_{bb}=U$ first. As shown in Fig.~\ref{Fig2}(a), there are three phases that exist in the phase diagram when the interexchange symmetry is preserved, namely 2SF, SCF, and 2MI. At $J=0$, the chemical potential width of all Mott lobes are 1 (in units of $U$), but they are separated from each other by an SCF phase with a chemical potential width $\mu=U_{ab}=0.5U$. Here, the SCF phase is such a state in which the order parameter $\langle ab^\dag\rangle$ is non-zero but the single component SF order $\langle a\rangle$ and $\langle b\rangle$ are both equal to zero. These order parameters and the averaged particle numbers as a function of $\mu$ are presented in Fig.~\ref{Fig2}(b) for a fixed value of $J$.  From Fig.~\ref{Fig2}(a)-(b), one can find that $\langle n_a \rangle$ and $\langle n_b \rangle$ are equal to the same integer in a single 2MI lobe because of the presence of interexchange symmetry. For example, $(\langle n_a \rangle, \langle n_b \rangle)=(1,1)$ and $(2,2)$ in the first and second 2MI lobe respectively.  In an SCF lobe, there exist particle fluctuations for single component bosons while $\langle n_{\rm{tot}}\rangle$ is still an integer constant. These results are well consistent with that in Ref. \cite{arXiv2021}.

On the other hand, a more complex phase diagram is observed when the interexchange symmetry is broken. We recall here that we break the interexchange symmetry of the model by setting $U_{bb}=0.8U \neq U_{aa}=U$. As Fig.~\ref{Fig2}(c) shows, there are two new phases compared with Fig.~\ref{Fig2}(a), namely SFb+MIa and SFa+MIb. In the SFb+MIa (SFa+MIb) phase, $b$ ($a$) component bosons are in a SF state and $a$  ($b$) component bosons form a MI state. Unlike the $U_{aa}=U_{bb}$ case, we find that there is no phase to separate the 2MI lobes at $J=0$ and they are located closely together. More importantly, the regions of the SCF phase are dramatically shrunk when the interexchange symmetry is broken. For example, there are two SCF lobes shown in Fig.~\ref{Fig2}(a), and then only the first SCF lobe is left in Fig.~\ref{Fig2}(c) when we set $U_{aa}\neq U_{bb}$. Those SCF lobes located in the large value of chemical potential are replaced by the 2MI phase when the interexchange symmetry is broken. In other words, it means that the prospects of observing an SCF phase in the interexchange symmetric case are much higher than the interexchange asymmetric case.

Besides, the averaged particle numbers per site of the interexchange asymmetric case are also plotted in Fig.~\ref{Fig2}(d) for a fixed value of $J$. Due to the imbalanced on-site intraspecies interactions, $\langle n_a \rangle$ and $\langle n_b \rangle$ need not be equal in a single 2MI lobe, e.g. $(\langle n_a \rangle, \langle n_b \rangle)=(1,2)$ in the second 2MI lobe. It also shows that the value of the total averaged particle number per site $\langle n_{\rm{tot}} \rangle=\langle n_{a}+n_{b} \rangle$ being jumped from an integer $k$ to the next integer $k+1$ between the two nearest-neighbor 2MI lobes. On the contrary, this value is changed discontinuously as $\mu$ goes from a 2MI lobe to the next 2MI lobe in the interexchange symmetric case, for instance,  $\langle n_{\rm{tot}} \rangle=2$ in the first 2MI lobe and then it is 4 instead of 3 in the second 2MI lobe as shown in Fig.~\ref{Fig2}(b).

\subsection{$W=-0.1$ case}
\begin{figure*}[htbp]
\centering
\includegraphics[width=0.9\textwidth]{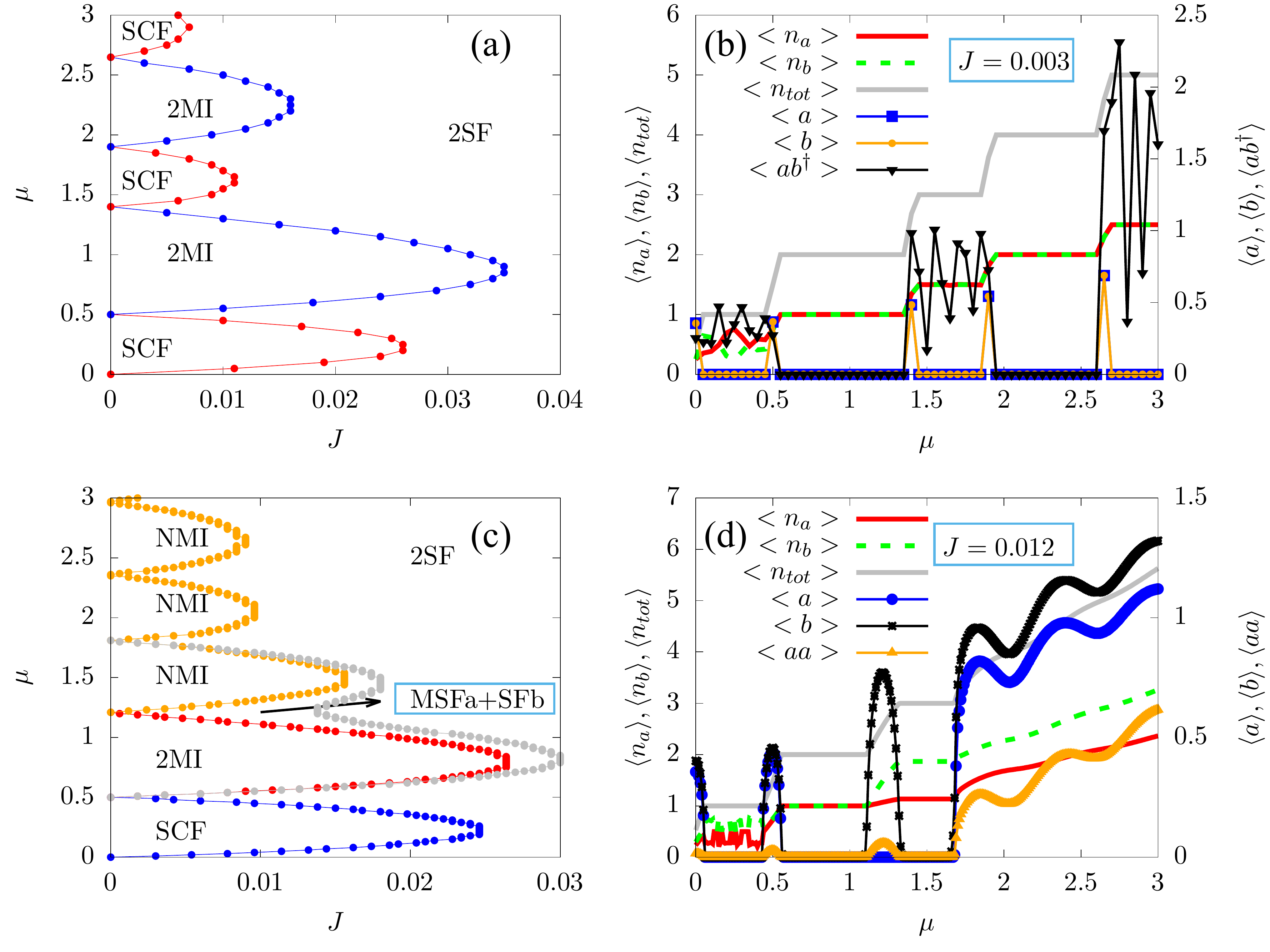}
\caption{\label{Fig3}(Color online) Phase diagrams, order parameters and particle distributions at $W=-0.1U$. Chemical potential $\mu$ versus $J$ (both are in units of $U$) phase diagram for (a) $U_{bb}=U_{aa}=U$ and (c) $U_{bb}=0.8U_{aa}=0.8U$, $U_{ab}$ is also set to $0.5U$ here as same as that in Fig.~\ref{Fig2}. Their corresponding averaged particle numbers and the order parameters as a function of $\mu$ are also shown in (b) and (d) respectively at a fixed $J$ value. In (b) and (d), the left (right) vertical axis represents the value of the particle number (superfluid order parameter). }
\end{figure*}
In order to study the influence of the pair hopping term on the phase diagram, the $W$ term is open in this section and the value is fixed at -0.1 (in units of $U$).  For the interexchange symmetric case, the phase diagram and the physical measurements are shown in Fig.\ref{Fig3}(a)-(b).  At $J=0$, the 2MI lobes are still separated from each other by an SCF state with a chemical potential width $\mu=U_{ab}=0.5U$. But instead of a fixed value in Fig.\ref{Fig2}(a), here the chemical potential width of the 2MI lobes at $J=0$ in Fig.\ref{Fig3}(a), $\Delta \mu^{\rm{2MI}}$, is a variable for different 2MI lobes, i.e. $\Delta \mu^{\rm{2MI}}$ is $0.9U$ and $0.75U$ in the first and the second 2MI lobe respectively. As $J$ increases from zero, the range of $\mu$ in 2MI and SCF state is decreasing, and these two states will disappear at a critical value of $J$, beyond which the system becomes a 2SF state. By comparing Fig.~\ref{Fig3}(a) and Fig.~\ref{Fig2}(a), when an interexchange symmetry is preserved, we surprisingly find that the $W$ term can not induce the new phase and can also not change the structure of the phase diagram.

Interestingly, when the interexchange symmetry is broken, two novel phases namely the MSFa and the NMI, which have been revealed in our previous work \cite{PhysRevLett125.245301}, are observed in the phase diagram. As Fig.~\ref{Fig3}(c) shows, the orange dotted lines correspond to phase boundary for the NMI state is determined from the non-integer value of $\langle n_{\alpha} \rangle$ ($\alpha=a,b$) and the zero single component boson SF order $\langle a \rangle=\langle b \rangle=0$, the gray dotted lines correspond to phase boundary for the MSFa state is determined from the non-zero molecular SF order of boson $a$ ($\langle aa \rangle \neq 0$) and zero atomic SF order of boson $a$ ($\langle a \rangle=0$).  In the MSFa phase, every two atoms of boson $a$ are paired together and form an SF state, while the $a$ atoms are in a MI state. In contrast with Fig.\ref{Fig3}(a), the Mott insulator lobes including 2MI and NMI in Fig.~\ref{Fig3}(c) are located closely at $J=0$.  As $J$ increases from zero, the range of $\mu$ in  NMI phase or MSFa phase is also decreasing, and the phases will disappear at a critical value of $J$.  From Fig.~\ref{Fig3}(a) and (c), we also learn that the region of the SCF phase in the phase diagram is still dramatically shrunk when the interexchange symmetry is broken. It is surprising that even if there has an SPH interaction,  two novel phases (NMI phase and MSFa phase) will not appear in the interexchange symmetric case.

In Fig.~\ref{Fig3}(d), we show the order parameters and the averaged particle numbers per site of the interexchange asymmetric case versus chemical potential at a fixed value of $J=0.012$.  In the NMI phase, although $\langle n_{\alpha} \rangle$ ($\alpha=a,b$) is an non-integer but $\langle n_{\rm{tot}} \rangle$ is still an integer, i.e. $\langle n_{\rm{tot}} \rangle =\langle n_{a} \rangle + \langle n_b \rangle=1.134174+1.865826=3$ for the first NMI lobe as shown by the integer-3 platform of the gray line in Fig.~\ref{Fig3}(d). The reason is that the total particle number per site is still a good quantum number when the $W$ term is turned on. Besides, the value of $\langle n_{\rm{tot}} \rangle$ being jumped from an integer $k$ to the next integer $k+1$ between the two nearest NMI lobes. While for the MSFa phase, the value of $\langle n_{\rm{tot}} \rangle$ is evolved with $\mu$ like that in a superfluid.

\subsection{Phase transition to the demixed states}
\begin{figure}[htbp]
\centering
\includegraphics[width=0.45\textwidth]{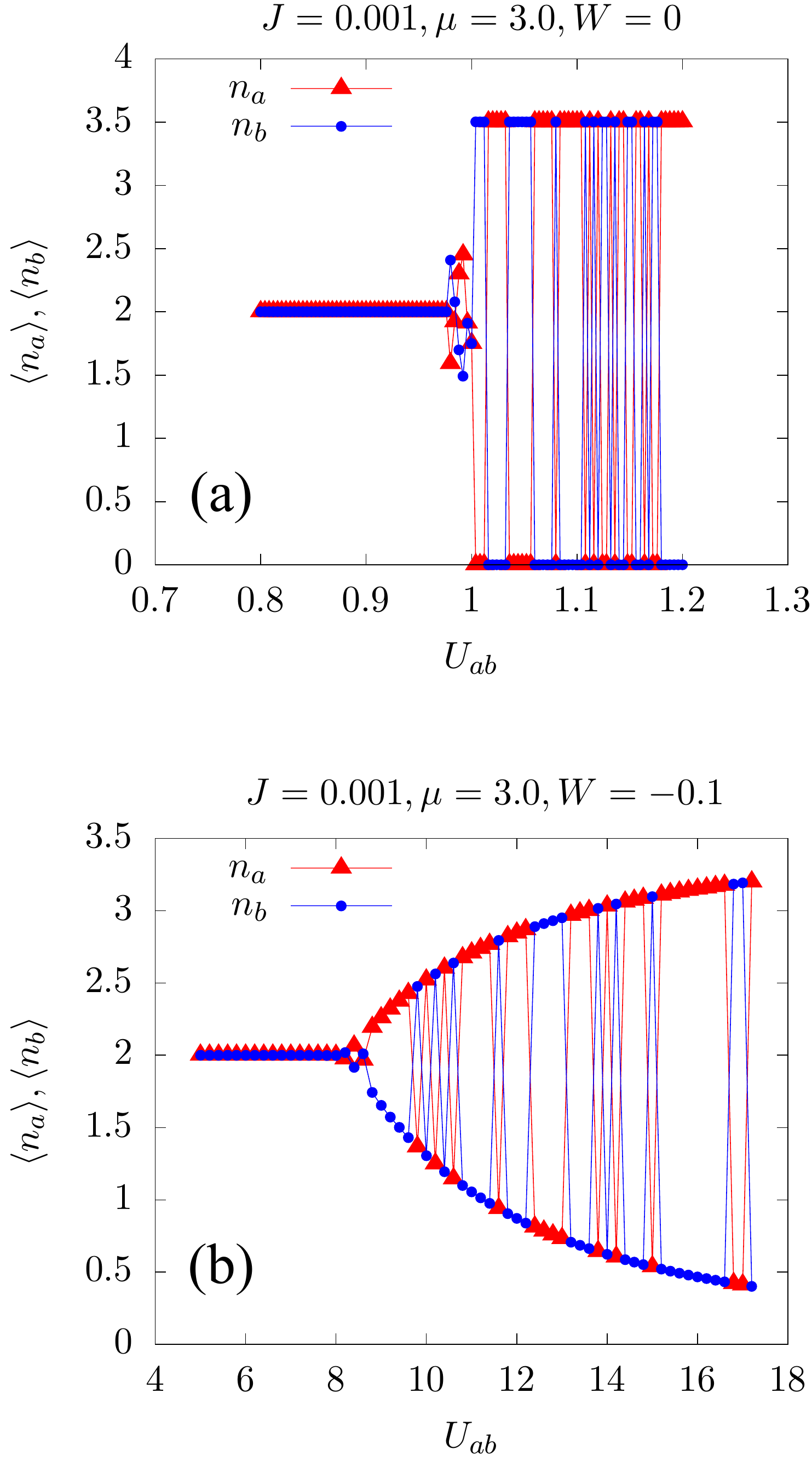}
\caption{\label{Fig4}(Color online) The averaged particle numbers per site as a function of $U_{ab}$ for (a) $W=0$ and (b) $W=-0.1$ (in units of $U$). Here, $U_{aa}=U_{bb}=U$ is set to be the energy unit and $(J,\mu)$ is fixed at $(0.001, 3.0)$.}
\end{figure}
To understand what has happened in the strong interspecies coupling limits, we calculate the value of $\langle n_a \rangle$ and $\langle n_b \rangle$ as a function of $U_{ab}$. For simplicity, we consider an interexchange symmetry here which requests that $U_{aa}=U_{bb}=U$ ($U$ is an energy unit). At $W=0$, people have known that the two-component bosonic system undergoes a phase transition to the demixed state when $U_{ab}>U$ \cite{PhysRevA76.013604, JPhysSocJpn81.024001, PhysRevA89.057601, PhysRevA92.053610}. We reproduce this result in Fig.~\ref{Fig4}(a) with $(J,\mu)$ is fixed at $(0.001,3.0)$. As it is shown in Fig.~\ref{Fig4}(a), the single species averaged particle numbers per site $\langle n_a \rangle$ and $\langle n_b \rangle$ are both equal to 2 at a small value of $U_{ab}$. It indicates that the system is in a 2MI phase and the two species are uniformly distributed in real space. Therefore, the 2MI phase represents a perfect mixed MI state.  On the other hand, if $U_{ab}>U$, as we have mentioned, the demixing effect occurs and the bosons from different species can not occupy the same site. Since only a two-site cluster (the remaining sites being replaced by a mean-field) is considered here, we expect that only one of the two species can be observed in the cluster when the system is in a demixed state because the two sites are equivalence. It means that one of $\langle n_a \rangle$ and $\langle n_b \rangle $ is non-zero and the other should be zero for $U_{ab}>U$. Therefore, we can distinguish this demixed phase through the value of $\langle n_a \rangle$ and $\langle n_b \rangle$, i.e. $\langle n_a \rangle=3.5$ and  $\langle n_b \rangle=0$ at a certain value of $U_{ab}$ ($U_{ab}$ is greater than $U$) as shown in Fig.~\ref{Fig4}(a).  Besides, it is completely random that a specific site is occupied by the $a$ or $b$ component bosons because of the interexchange symmetry. It demonstrates that whether the value of $\langle n_{\alpha} \rangle$ ($\alpha=a,b$) is zero or non-zero should be random for $U_{ab}>U$.  All these features can be seen in Fig.~\ref{Fig4}(a). We also notice that in this demixed state, each species is in an SF phase owing to the non-integer value of $\langle n_{\alpha} \rangle $($\alpha=a$ or $b$), but they are separated in real space and implying a demixed SF phase is formed.

Let us now examine how the pair hopping term affects the demixing effect. We turn on the pair hopping term and $W$ is set to be $-0.1U$. The results are presented in Fig.~\ref{Fig4}(b). By contrast, the mixed-demixed transition point is sharply shifted to a large value of $U_{ab}$. More specifically, here the critical point is $U_{ab}\approx 8U$ instead of $U_{ab}=U$. Moreover, the demixing effect for this case is not complete due to the mechanism that the $W$ term can always lower the energy of the system via the on-site particle exchange between the two species. Therefore, either of $\langle n_a \rangle$ and $\langle n_b \rangle$ should not be exactly equal to zero at finite $U_{ab}$ in the demixed state, while one of them is very small and tend to be zero when $U_{ab} \rightarrow \infty$. This incompletely demixed state is still a demixed SF state because of the non-integer averaged particle number.

\section{Conclusions}\label{Conclusions}
We have investigated the ground-state phase diagram and the demixing effect of a two-component bosonic system with pair hopping in synthetic dimension via the cluster Gutzwiller mean-field method. We find that the region of the SCF phase is dramatically shrunk in the phase diagram when the interexchange symmetry is broken for both $W=0$ and $W=-0.1U$ cases, i.e. there is only one SCF lobe shown in Fig.~\ref{Fig2}(c) and Fig.~\ref{Fig3}(c). Therefore, our results reveal that the prospects of observing the SCF phase are much higher in the interexchange symmetric case. Unexpectedly, the NMI phase and the MSFa phase can only be observed in such a two-component bosonic system that the pair hopping term is opened and its interexchange symmetry should be broken. More notable, the structure of the phase diagram can only be changed by including the SPH interaction and simultaneously breaking interexchange symmetry. Furthermore, the demixing effect in a two-component bosonic system with SPH interaction is also revealed. The system undergoes a phase transition to the demixed phase when the strength of $U_{ab}$ is sufficiently large. At $W=0$, this demixing effect is perfect, i.e. the bosons from the different species can not occupy the same site absolutely when $U_{ab}>U$. While for a finite value of $W$, there always exists particle exchange between the two species on the same site, and thus an incompletely demixed state is formed beyond its critical point. We hope all these predictions can be examined in future experiments.

\section{Acknowledgements}\label{Acknowledgements}
The authors thank Y. Chen and  J. Lou for fruitful discussions.  This research is supported by the National Natural Science Foundation of China (NSFC) under Grant Nos.11947102 and 12004005, the Natural Science Foundation of Anhui Province under Grant No. 2008085QA26, and the Ph.D. research Startup Foundation of Wenzhou University under Grant No. KZ214001P05, and the open project of state key laboratory of surface physics in Fudan University (Grant No. KF2021$\_$08).

\end{document}